\documentclass[aip,rsi,amsmath,amssymb,floatfix,superscriptaddress,%
showkeys,%
reprint,%
author-numerical,%
]{revtex4-1}

\usepackage{graphicx}% Include Fig. files
\usepackage{dcolumn}% Align table columns on decimal point
\usepackage{bm}% bold math
\usepackage{}
\usepackage[T1]{fontenc}
\usepackage{mathptmx}
\usepackage{textcomp}
\usepackage{multirow}
\usepackage{xcolor}
\usepackage{enumitem}

\usepackage{xr}
\usepackage{soul}
\usepackage{xr}
\makeatletter
\newcommand*{\addFileDependency}[1]{% argument=file name and extension
  \typeout{(#1)}
  \@addtofilelist{#1}
  \IfFileExists{#1}{}{\typeout{No file #1.}}
}
\makeatother

\newcommand*{\myexternaldocument}[1]{%
    \externaldocument{#1}%
    \addFileDependency{#1.tex}%
    \addFileDependency{#1.aux}%
}
\myexternaldocument{SupInfo}

\begin{document}
\title{Real-time visualization of metastable charge regulation pathways in molecularly confined slit geometries}

\author{H.-W. Cheng}
\email{hsiu-wei@iap.tuwien.ac.at}
\affiliation{Institute of Applied Physics, Vienna Institute of Technology, Wiedner Hauptstrasse 8-10/E134, 1040 Wien, Austria}

\author{J. Dziadkowiec}
%\email{dziadkowiec@iap.tuwien.ac.at}
\affiliation{Institute of Applied Physics, Vienna Institute of Technology, Wiedner Hauptstrasse 8-10/E134, 1040 Wien, Austria}
\affiliation{NJORD Centre, Department of Physics, University of Oslo, PO Box 1048, Oslo, Norway}

\author{V. Wieser}
%\email{valentina.wieser@tuwien.ac.at}
\affiliation{Institute of Applied Physics, Vienna Institute of Technology, Wiedner Hauptstrasse 8-10/E134, 1040 Wien, Austria}

\author{A. M. Imre}
%\email{(imre@iap.tuwien.ac.at) }
\affiliation{Institute of Applied Physics, Vienna Institute of Technology, Wiedner Hauptstrasse 8-10/E134, 1040 Wien, Austria}

\author{M. Valtiner}
\email{valtiner@iap.tuwien.ac.at}
\affiliation{Institute of Applied Physics, Vienna Institute of Technology, Wiedner Hauptstrasse 8-10/E134, 1040 Wien, Austria}

\date{\today}

\begin{abstract}

Transport of ions in molecular-scale confined spaces is central to all aspects of life and technology: into a crack, it may break steel within days; through a membrane separator, it determines the efficiency of electrochemical energy conversion devices; or through lipid membranes, it steers neural communication. 
Yet, the direct observation of ion mobility and structuring in sub-nanometer confinement is experimentally challenging and, so far, solely accessible to molecular simulations. 
Here, we show quantitative, 3D molecularly-resolved ion transportation of aqueous ionic liquid and s-block metal ion solutions, confined to electrochemically-modulated, molecular-sized slits.  
Our analysis of atomically resolved solid/liquid interface unveils generic rules of how enthalpic ion-ion and ion-surface interactions and entropic confinement effects determine the charge regulation mechanism. 
Altering our general understanding, the confined charge regulation may proceed \textit{via} fast, kinetically favoured, metastable pathways, followed by slow diffusive thermodynamic ion reorganization, which has important implications for all charge-regulated systems. 
\end{abstract}

\maketitle
\begin{figure*}[t]
    \centering
    \includegraphics[scale=0.95]{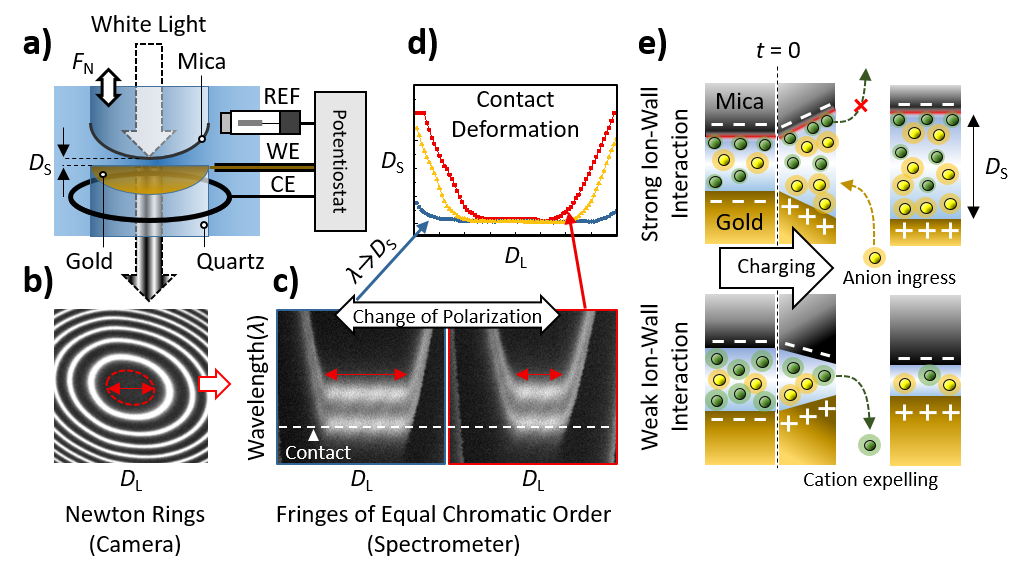}
    \caption{\textbf{a)} Schematic representation of the 3-electrode electrochemical sSFA setup: A (mica) template-stripped, semi-transparent (40 nm thick) gold layer on a cylindrically curved quartz disk comprises working electrode (WE); platinum ring acts as a counter electrode (CE); and a Ag|AgCl mini-electrode is a reference electrode (REF). 
    Cleaved, thin mica surface (< 5 $\mu$m), back-coated with a semi-transparent (40 nm thick) silver layer, and the gold film form an optical cavity, which produces an intererometric pattern on passing the white light through the two contacting surfaces. 
    \textbf{b)} The resultant interferometric pattern reveals Newton rings that connect the regions of the same surface separation. The most central Newton ring outlines a spherical contact region (indicated with a red dashed circle), where \textit{D}$_L$ is the lateral distance. 
    \textbf{c)} The transmitted discrete light (standing wave) is split in the spectrometer as fringes of equal chromatic order (FECO). FECO provide the information about the separation distance (\textit{D}$_S$) between the two surfaces acting as semi-transparent mirrors. The flattened part of FECO indicate the diameter of a round-shaped contact between the two surfaces (indicated by red arrows) and the overall curved profile outlines the contour of a gradually opening gap. 
    \textbf{d)} Varying contact deformation reconstructed by fitting the changing separation distance from the recorded FECO patterns upon a change in the surface polarization. 
    \textbf{e)} Cartoon showing possible, electrochemically-driven charge regulation pathways in a molecularly confined mica-gold slit, with strong or weak ion-surface interactions. The charge regulation mechanisms, driven by the modulation of the externally applied surface potential, E\textsubscript{WE}, can be resolved in real-time and with a molecular-scale resolution by measuring the separation distance \textit{D}$_S$ and normal forces \textit{F}$_N$ between two apposing surfaces in the electrochemical sSFA.}
    \label{fig1r}
\end{figure*}

\section{Introduction}
Charge regulation kinetics within molecular sized pores or compartments control central aspects of our perceived macroscopic world\cite{ChengValtiner2020}.
For instance, in biological systems fluids are confined to less than 50 nm within any point of a cell\cite{gershon1985cytoplasmic}. 
This marked confinement drives complex molecular reaction kinetics that underlie biological life and regulate biological energy conversion within compartmentalized, electrochemical respiratory chains. 
Similarly, confinement effects become essential when scaling down energy storage and conversion devices such as fuel cells, batteries or electrocatalysts. 
Their increasing efficiency requires nanocomposite-based electrodes \cite{pean2015confinement, ling2014flexible}, higher energy densities \cite{kim2009high, tang2013ultra}, and tunable discharging times\cite{song2012polymer, zhang2016giant}. 
However, nano-structuring dramatically alters the charge/mass transport mechanisms\cite{Galeano2012Dec, ChengValtiner2020}.

Also in nano-fluidics, stress-corrosion cracking, or geological dissolution processes, fluid-filled nano-channel geometries substantially alter ion \textit{diffusion} (concentration gradient driven) and \textit{migration} (field gradient driven) and the locally acting osmotic pressures, unlike in systems with limited spatial confinement \cite{das2011steric,merola2017situ,kristiansen2011pressure,dziadkowiec2019nucleation}.
Thus, experimental measurements of ion diffusion kinetics during a surface charging/discharging process in nano-cavities are essential to progress our understanding of charge regulation across all those systems. 

Electrochemical potential shifts, applied to walls of a nano-confined (20 - 500 nm) slit in the interferometry-based Surface Forces Apparatus (SFA) technique, drive recurring changes both in slit separation and in the measured interaction forces.\cite{shrestha2014angstrom,tivony2021modulating}
These transient forces correlate with the ongoing charge regulation inside an electrochemically reactive nano-slit.
Also, for non-reactive surface charging\cite{tivony2018charging}, the slower ion migration in 20-50 nm gaps is quantitatively connected to the degree of spatial confinement. 
Ion-to-pore size ratios\cite{largeot2008relation, lin2009solvent}, and ion-surface interactions under polarization\cite{baldelli2008surface,nishi2013ultraslow,lauw2012structure}, play a key role during such nano-pore charging processes.

Previous works have quantified 2D ion diffusion through Ångstrom-thick slits at the molecular level\cite{esfandiar2017size}. 
However, a molecularly-resolved visualization of ion migration kinetics in 3D subnanometer confinement has been so far solely reconstructed from simulation studies\cite{gupta2020charging}, indicating a decisive effect of ion-surface interactions\cite{kondrat2014accelerating}.
The ion-specific and surface chemistry-sensitive strength of counterion adsorption onto confining walls with variable surface charge may be used to tune the electrically-induced flows in extremely thin nano-fluidic devices, where the electric double layers (EDLs) are highly overlapped, having implications for nano-filtration, membrane-based concentrators, and flow generators in molecular sensing applications \cite{chang2009understanding,mani2009propagation,rubinstein2008direct,kim2007concentration}.
 
Here we use a novel electrochemically modulated \textit{in situ}-sensing Surface Forces Apparatus (sSFA)\cite{Wieser2021} to resolve molecularly both the interfacial ion structuring in the inner double layer and the ion exchange kinetics in a slit formed between two cross-cylindrical disks, with the separation distance between the two confining walls \textit{D}$_S$ of~0.5-8.5 nm. 
Figures \ref{fig1r}\textbf{a)} through \ref{fig1r}\textbf{d)} demonstrate how the nanometer slit is generated and interferometrically measured/visualized using the sSFA.

In such extreme confinement, where the distance $D_S$ between the apposing surfaces is < 2-3 nm, the natural or applied charging of the surfaces determines the resulting ion configuration in the gap, where charge neutrality is a boundary condition. 
As gaps are only molecularly wide, such configurations essentially arise from the overlap of two interacting EDL (with specifically and non-specifically adsorbed counterions at the inner and outer Helmholtz planes, as described in the EDL Stern model \cite{israelachvili2011intermolecular}). 
Fig. \textbf{\ref{fig1r}e)} illustrates the possible ionic responses during charge regulation in a molecularly confined nano-slit, which recover charge neutrality in the gap by means of expelling or refilling of cations and anions. 

In molecular confinement, charge regulation may proceed through various possible kinetic paths towards a thermodynamic equilibrium. 
In particular, strong ion-ion and ion-surface interactions become significant in altering these kinetic pathways\cite{rotenberg2015structural}.

In this study, we examine these interactions for water soluble chloride-based ionic liquids (IL) (C$_2$MImCl, C$_8$MImCl) and inorganic electrolytes (LiCl, NaCl, CsCl, CaCl$_2$) confined between an electrochemically modulated gold surface and a negatively charged mica substrate.
Both the varying hydrophobicity of ILs (tail-tail interactions) and varying hydration energies of the (earth) alkali cations moderate ion-ion and ion-surface interactions. 
Using this model system, we are able to trace the real-time ionic responses with molecular precision.

\section{Results and Discussion}

\begin{figure}[t]
    \centering
    \includegraphics[scale=0.6]{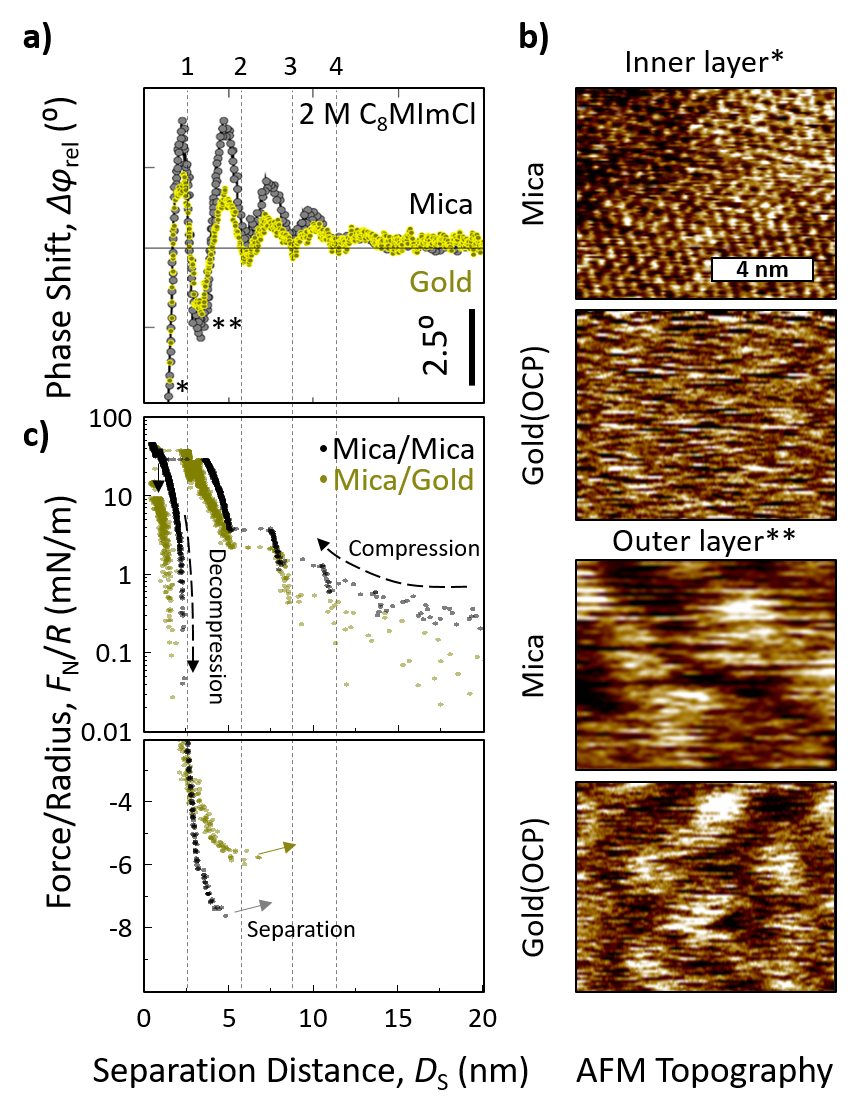}
    \caption{Structuring of C$_8$MImCl on gold and mica surfaces in 2 M solution:
    \textbf{a)} Amplitude Modulated AFM (AM-AFM) phase-distance profiles for mica-IL (grey) and gold-IL (yellow) interfaces in 2 M solutions; AFM topographies show \textbf{b)} structurally resolved images, acquired at a selected setpoint of the outer (**) or inner (*) layers marked in \textbf{b)}; \textbf{c)} SFA Force-Distance (F-D) curves of symmetric mica-mica (black) and asymmetric mica-gold (yellow) systems; for the mica-mica experiment more details are shown in the SI (Figure \textbf{S2})).}
    \label{fig2r}
\end{figure}

\begin{figure}[t]
    \centering
    \includegraphics[scale=0.55]{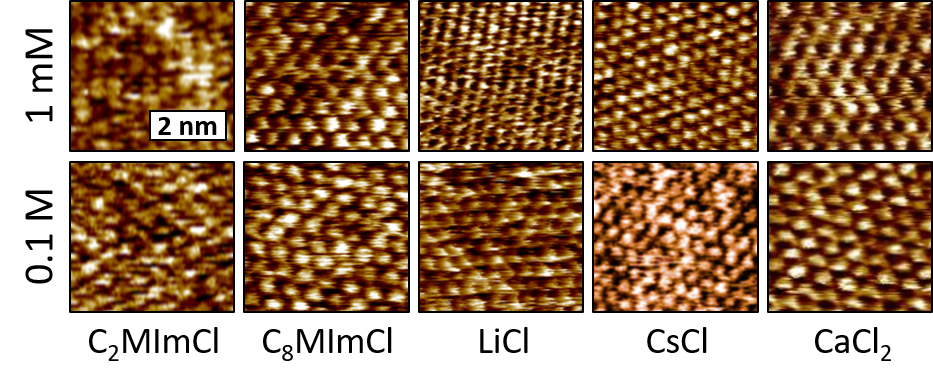}
    \caption{Atomically resolved, concentration-dependent ionic structure topography on mica surface for all salts used in this work acquired in AM-AFM.}
    \label{fig3r}
\end{figure}

Throughout this work, we draw a comprehensive picture of the charge regulation pathways in relation to the ionic interfacial structuring by comparing the behaviour of IL solutions and (earth) alkali salt solutions.

We first used the amplitude-modulated AFM (AM-AFM) to examine the differences in interfacial ion structuring. 
Fig. \textbf{\ref{fig2r}} compares \textbf{a)} phase-distance profiles and \textbf{b)} topography of C$_8$MImCl ion structuring at given setpoints, in 2 M solutions both on mica and gold (at open circuit potential; OCP).

Oscillatory phase-distance profiles were observed in AM-AFM for C$_8$MImCl both on mica (grey) and on gold (yellow). 
When overlapped, the phase-distance profiles show the same periodicity on gold and mica, but the magnitude of phase oscillations is reproducibly weaker on gold. 
This suggests a less ordered IL cation structure on gold, evidenced previously for other moderately charged surfaces\cite{bou2010nanoconfined}.

AM-AFM imaging of the Helmholtz layers of mica and gold (marked by *) in C$_8$ImCl solutions are shown in Fig. \textbf{{\ref{fig2r}b)}}. This data revealed a molecularly-resolved, well-ordered IL cation structure on mica.
No ordered IL structure was resolved on gold at the same setpoint, confirming weak ion-gold interactions.
Imaging of the second, outer layer on both surfaces (marked by **) exposed patterns that were very distinct from the inner layers with no pronounced molecular ordering. 
Instead, similar undulating topographies were revealed on both surfaces. 
Such patterns indicate the presence of lamellar IL bilayers on both substrates.

The AFM images shown in Fig. \textbf{{\ref{fig2r}b)}} are statistically relevant and selected based on the analysis of a series of phase and height AM-AFM images from low (close to the inner layer) to high (outside of the outer layer) setpoint on mica. 
Specifically, the statistical image analysis of phase shifts and standard deviations follows the oscillatory profile, providing a clear selection rule for images of given layers (details are given in Fig. \textbf{S1}).
This image analysis confirms the presence of lamellar layering and the resultant phase shift variations on a global scale.

The correspondingly selected AFM topographies of mica in 1 mM and 0.1 M C$_2$MImCl, C$_8$MImCl and other (earth)alkali chloride solutions are shown in Fig. \textbf{\ref{fig3r}}.
We found pronounced ion structuring and thus strong mica-ion interactions for CsCl and CaCl$_2$, while the topographies measured for LiCl and C$_2$MImCl indicate disordered, diffuse Helmholtz Layer.

In general, none of the cations has been reported to undergo ordered surface adsorption on gold; conversely, it is well established that gold-adsorbed chloride ions form a monolayer at positive potentials above 600 mV vs. Ag|AgCl\cite{shi1996chloride, higashi2015two}.

We accessed additional details on the IL structuring with the sSFA force measurements. Fig. \textbf{\ref{fig2r}c)} compares C$_8$MImCl system probed in the symmetric mica vs. mica and asymmetric gold vs. mica configurations. 
The concentration-dependent force vs. distance (F-D) characteristics and layer-by-layer approach curves are also additionally shown in the SI (Fig. \textbf{S2}, \textbf{S3}). 
Unlike AFM, which pokes through the interface-adsorbed molecular layers with a very sharp tip, sSFA is compressing the confined solution in a micron-wide confinement and expels entire molecular layers at a characteristic load threshold (see Fig. \textbf{S2c)}).

Our results show that molecular layers measured in sSFA displayed an oscillatory profile of 2.2 nm, which matches the AFM phase-distance characteristics, revealing the multilayered IL structures with complementary techniques\cite{cheng2018effect}.
Comparably weaker but still clearly oscillatory sSFA F-D profiles (less load required to expel a given molecular layer) were resolved in the IL-gold system.

This again confirms that the molecular structure on gold was less ordered than on mica.
Such poor ordering can be attributed to the weaker electrostatic interactions between IL cations and the gold surface because of the low surface charge at the OCP (ca. 0.2-0.4 V).
At low concentrations this behaviour was also confirmed by diffuse double layer potentials fitted using the DLVO theory (see SI, Table \textbf{1} and Fig. \textbf{S3b)}).
These show that gold displayed a charge regulated behaviour during the EDL overlap, while mica exhibited a constant charge behaviour, which points to the weaker ion-gold interactions and relatively strong ion-mica interactions.

In contrast to C$_8$MImCl, C$_2$MImCl appeared more diffuse and exhibited molecular vacancies in the atomically visualized surface layer (see Fig. \textbf{\ref{fig3r}}) at 1 mM and at 0.1 M.
The cations were reproducibly less ordered, which is indicative of weaker ion-surface interaction. 
As a result, in comparison to C$_8$MImCl, no pronounced oscillatory but diffuse force profiles were observed in the sSFA data of C$_2$MImCl in the mica-mica configuration (see again Fig. \textbf{S2a)}). 
Similarly, AFM topography of LiCl showed Helmholtz layer structure with a diffuse double layer, while the imaging of mica in CsCl and CaCl$_2$ solutions showed distinct and well-ordered cation layering, as reported in previous works\cite{baimpos2014effect}. 

Hence, we can conclude that molecular ordering on the isolated surfaces (which is a direct result of ion-surface interaction strengths) decisively controls the characteristics of F-D profiles. 
Weak interactions and disordered structures result in the diffuse EDL overlap with less pronounced oscillatory behaviour and vice versa. 

\begin{figure*}[!t]
\includegraphics[scale=0.6]{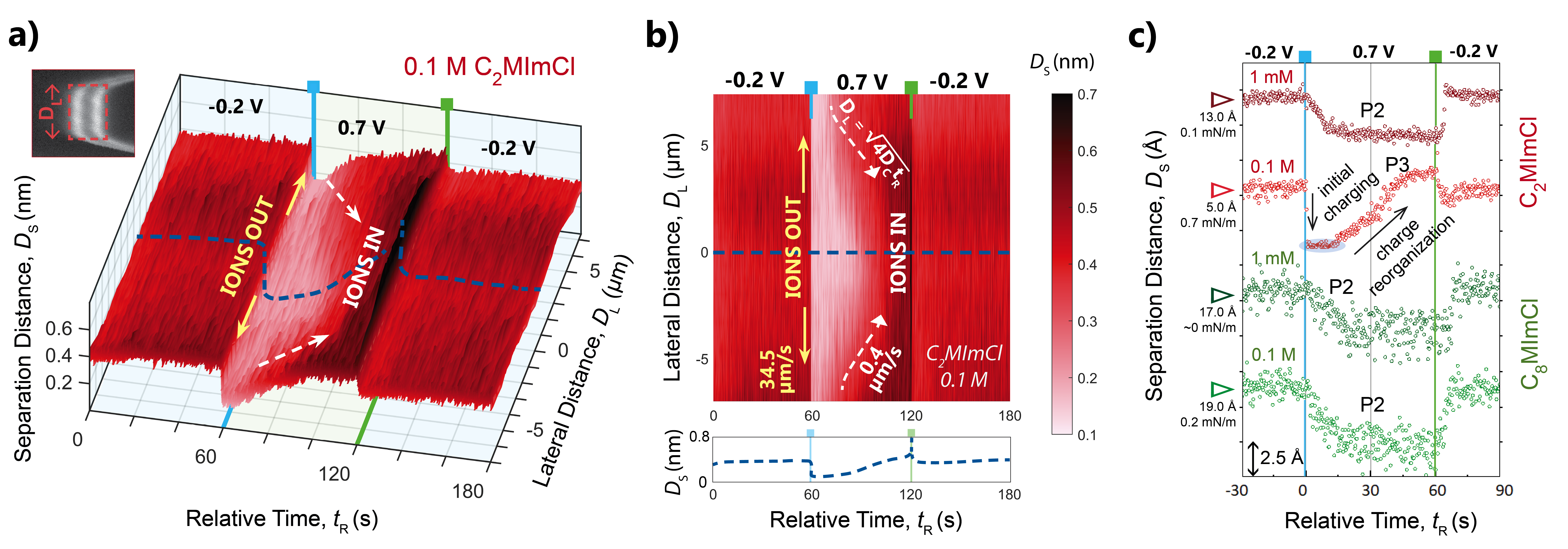}
    \caption{Electrochemically-modulated charging within the mica-gold nano-slit measured in ionic liquid solutions using sSFA: \textbf{a}) 3D map showing 0.1 M C$_2$MImCl diffusion during slit charging at the applied potentials of -0.2~V and +0.7~V. 
    The gap thickness variation (separation distance; D$_S$) in response to the potential switches is shown as a function of time across the whole flattened contact region (lateral distance; D$_L$).
    The flattened contact region is outlined in the FECO inset in the top left corner.
    The dashed blue line marks the separation distance change in the centre of the contact region, used to plot the separation distance line profiles throughout this work; 
    \textbf{b)} 0.1 M C$_2$MImCl projection view of panel \textbf{a)} showing the diffusion profile during slit charging. 
    The separation distance, D$_S$, is indicated with a color bar. 
    The map highlights the fast removal of confined ions upon the potential switch from -0.2~V to +0.7~V followed by the slower diffusion of ions back into the gap; 
    \textbf{c)} The separation distance of centre line profiles during polarization switches are compared for C$_2$MImCl and the more hydrophobic C$_8$MImCl solutions, both at 1 mM and 0.1 M concentrations showing fast initial charging.
    In the case of C$_2$MImCl we observed an additional slower charge reorganization behavior. The initial separation distance and applied load are indicated for each profile.}
    \label{fig4r}
\end{figure*}

\begin{figure*}[!t]
\includegraphics[scale=0.5]{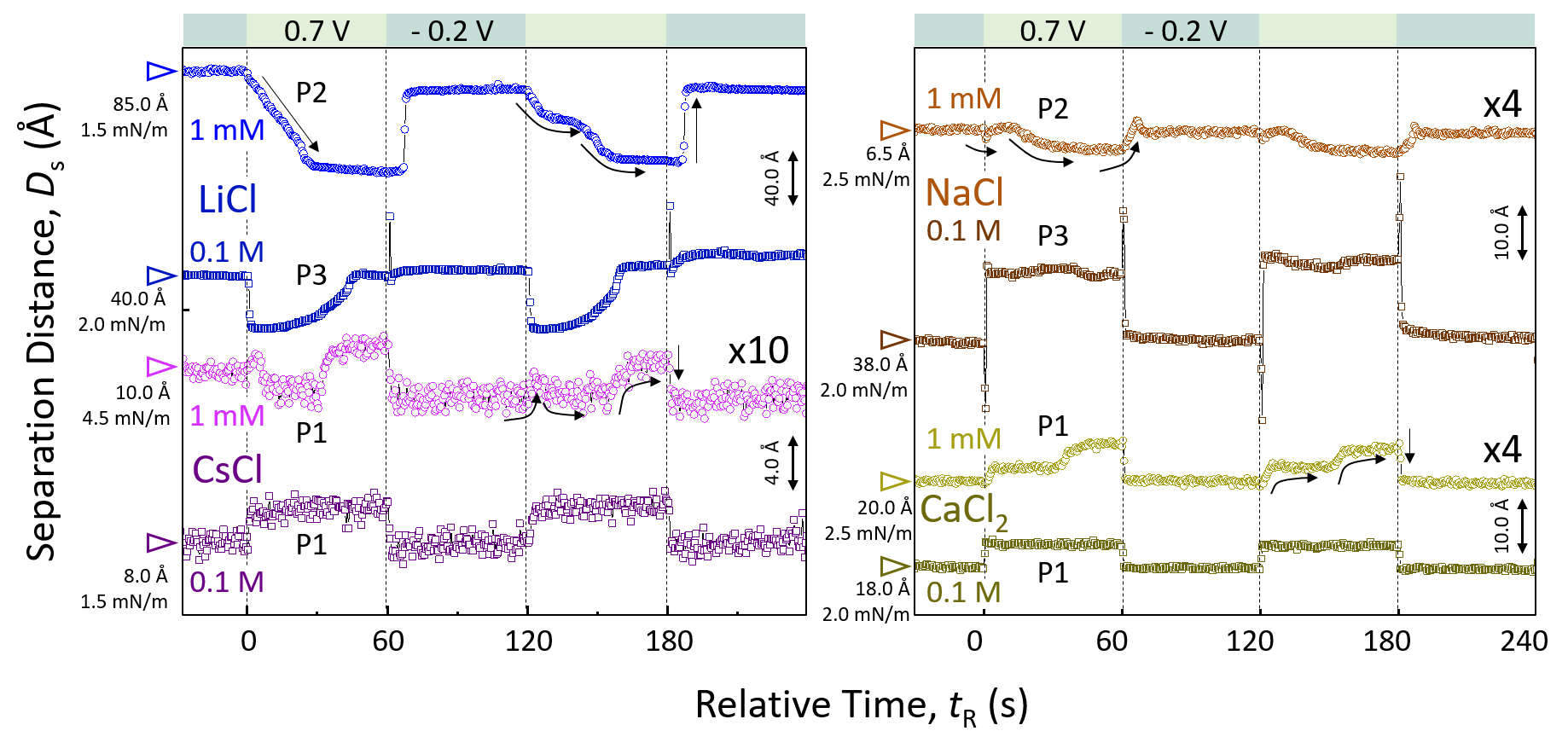}
   \caption{Electrochemically-modulated charge regulation within a mica-gold nano-slit in 1 mM and 0.1 M aqueous chloride solutions: Gap thickness change line profiles measured in the centre of confinement in response to the square wave potential change between -0.2~V (blue shading) and +0.7~V (green shading) for \textbf{a)} LiCl and CsCl and \textbf{b)} for NaCl and CaCl$_2$. The absolute separation at the start of polarization cycles and the initial applied load is given for each experiment. Note the varying separation distance scale for different salts as marked with respect to LiCl and with the individual scale bar for each salt.}
   \label{fig5r}
\end{figure*}

Based on this molecular-level understanding of the static EDL structuring we can now move on and discuss the electrochemically triggered ionic responses in confinement between a polarizable gold surface and a permanently negatively charged mica surface in our sSFA setup (see again Fig. \textbf{\ref{fig1r}a)}).

Fig. \textbf{\ref{fig4r} a, b)} show the 3D gap thickness profiles of the flattened mica-gold contact region in response to periodic electrochemical square-wave polarization (-0.2~V and +0.7~V, 8.33 mHz) for 0.1 M C$_2$MImCl ionic liquid solution.  
This 3D tracking of the gap separation over time across a large, micron-scale lateral confinement revealed the abrupt ion removal upon the initial charging from -0.2~V to +0.7~V, followed by the diffusive inward migration of ions back to the confined slit. 
On the reverse polarization, the ionic composition of the gap at -0.2~V was restored. 
For simplicity, the slit charging profiles are further shown as the gap thickness profiles in the center of confinement, as indicated with blue dashed lines in Fig. \textbf{\ref{fig4r} a, b)}.

The center line profiles showing the change of the separation distance induced by the gold polarization switches for C$_2$MImCl and C$_8$MImCl ionic liquids are compared in more detail in Fig. \textbf{\ref{fig4r} c)}. 
Interestingly, the gap thickness response for 0.1 M C$_2$MImCl solution varied significantly from those measured at lower concentration, and compared to the one for C$_8$MImCl. As marked, after the initial charging a second, slower charge reorganisation process occurred. 

Reproducible gap charging profiles measured for (earth) alkali electrolytes at 1 mM and 0.1 M concentrations are further shown in Fig. \textbf{\ref{fig5r}}.
These ions also displayed a range of intricate ion-specific diffusion patterns as discussed below (for more details also refer to Fig. \textbf{S4} and Fig. \textbf{S5} for the full 3D pore charging profiles). The control measurement of gap charging in pure MilliQ water is also shown in Fig. \textbf{S6}. 

Based on the separation distance line profiles in the centre of confinement, we observed three distinct ionic responses (referred to as patterns; P1-3) induced by the anodic (from -0.2~V to +0.7~V) polarization of the gold electrode:\\
P1) slit swelling with no observable cation expelling and sole anion refilling at both low and high concentrations (CsCl, CaCl$_2$; as seen in Fig. \textbf{\ref{fig5r}});\\
P2) slit thickness decrease with slow cation expelling for C$_2$MImCl, LiCl and NaCl at 1 mM, and for C$_8$MImCl at both concentrations.\\
P3) fast cation expelling, which drives the system into a transient state, followed by a surprising subsequent concentration-dependent refilling against the entropic penalty observed for C$_2$MImCl, LiCl and NaCl at 0.1 M (see Fig. \textbf{\ref{fig4r}c)} and Fig. \textbf{\ref{fig5r}});\\

P2 and P3 occurred in the absence of strong ion-mica interactions (diffuse EDL), while P1 was observed for systems with strong binding of cations on the basal planes of mica (see AFM images in Fig. \textbf{\ref{fig3r}}). These results agree with the thermodynamic trend of cation-mica interaction\cite{jia2018reversal}, where the adsorption tendency is Ca$^{2+}$ $\sim$ Cs$^+$ >> Na$^+$ $\sim$ Li$^+$. 

P2 and P3 revealed concentration-dependent behaviour for poorly-adsorbing C$_2$MIm$^+$, Li$^+$, and Na$^+$ cations. 
In P2, at low 1 mM concentration of inorganic cations, this slow cation removal was related to the lower concentration gradients between the confined slit and the bulk solution at 1 mM, which resulted in slower exchange rates, i.e. slower kinetics. At higher 0.1 M concentration, the exchange kinetics for these cations was considerably faster (P3). 
In contrast, for C$_8$MImCl even at 0.1 M, we always observed a slow ion ejection (P2). 
Here, this can be attributed to the pronounced lateral hydrophobic interaction of the cations with long hydrocarbon tails (which is effectively similar to jamming\cite{israelachvili2011intermolecular}); see Fig. \textbf{\ref{fig4r}c)}.

Although these cations were removed from the slit on anodic polarization both at low and high concentrations, this process became faster and more complex at higher 0.1 M concentration: The initial fast cation expelling in P3 was induced by the charge regulation, later followed by the slit relaxation and swelling due to ion pairs reentering the confined region.

Thus, as especially evident for C$_2$MIm$^+$ and Li$^+$ at higher concentrations, P3 corresponds to a two-step kinetically-controlled path toward a thermodynamic equilibrium as follows:
During step 1, the initial fast decrease in the slit thickness originated from the expected migration of cations, which occurred in response to the instantaneous increase of the positive charge density on gold. 

Step 2 instead comprised a diffusive (and hence slower) charge reorganization process towards the thermodynamic equilibrium, which involved ion pairs refilling the confined space under the charge neutrality condition. This second step is driven by enthalpic energy gain due to the strong gold-chloride interaction, as ion confinement is clearly entropically unfavourable.
Analogously transient ionic reconfiguration occurred significantly faster in 0.1 M NaCl system. 
The lack of gap swelling and ion pair refilling at +0.7~V (step 2) for 1 mM  C$_2$MImCl, LiCl and NaCl  (P2) indicates a metastable, cation-depleted state. Here, the ions reassembled more slowly, as determined by the concentration gradients between the bulk and the confined region, which control the diffusive motion of charges.

In contrast, as seen in P1, Ca$^{2+}$, and Cs$^+$ (which are surface potential determining cations for mica \cite{baimpos2014effect,kekicheff1993charge}) remained bound within the gap.
Here, the enthalpic ion-surface interaction overcame the entropy-favoured ion discharge.  
As such, the charge neutrality had to be restored by extra anions migrating in from the bulk. Such charge regulation for ions strongly adsorbed onto mica is well illustrated with the CsCl example:
While at high 0.1 M CsCl concentration, the slit was charge regulated simply by chloride ingress upon anodic polarization, the charge regulation at 1 mM took an intricate detour. First, with a minute initial chloride influx (increase of $D_S$), then followed by slow cation ejection (decrease of $D_S$), and with final thermodynamic stabilization at larger $D_S$.
The initial Cl$^-$ migration in this case breaks down due to the ion depletion at the gap opening. 
As a result, the inflow of chlorides is rate-limited by the diffusion from the bulk, while slow outwards migration of Cs$^+$ (due to strong ion-surface interaction) proceeds towards a metastable charge neutrality.

Thus, from a thermodynamic standpoint the charge regulation process in the confined sSFA geometry upon anodic polarization is generally driven by a balance between entropic and enthalpic effects. In the case of weak ion-surface interaction (diffuse EDL on mica and/or gold), entropy may favour depletion of cations from confinement, which leads to a decrease of separation distance between the confining walls.
However, if strong ion-surface interactions are present, cations will remain in the slit, which forces anions to flow in to reestablish the charge neutrality in confinement, thereby swelling the slit.
Although entropically unfavourable, this second scenario is driven by an enthalpic (electrostatic and specific surface) energy gain\cite{shi1996chloride, higashi2015two}.

The reverse polarization (+0.7~V to -0.2~V) generally restored the initial gap thickness \textit{via} one step ion migration, inducing a jump to the initial equilibrium gap thickness. 
For 0.1 M Na$^+$ and Li$^+$, an additional fast slit swelling was observed prior to the jump. 
This is possibly related to the rehydration of chloride ions. 
Chlorides specifically adsorb to gold, and need to hydrate before they are expelled, giving rise to the initial swift increase in separation. 
Further, although Cl$^-$ are known to over-adsorb as a monolayer on gold at positive polarizations\cite{shi1996chloride}, Cl$^-$ layers can be easily removed from the charge-regulated gold surface, shrinking the gap for CsCl, NaCl and CaCl$_2$ electrolytes on the reverse polarization. For the ILs (with the exception of 0.1 M C$_2$MImCl), the cation migration inwards is the direct path to charge neutrality because of the lack of Cl- anions in the cation-dominated gap.

In addition to the ion diffusion tracking at the center line of confinement, Fig. \textbf{\ref{fig4r}b)} shows a time series that directly visualizes 3D transportation of C$_2$MImCl ions across the whole confined zone (others are shown in Fig. \textbf{S4}).
By tracing the moving, monolayer-thick front of the gap thickness change, we directly measure the flow rates of the ionic charge regulation waves as indicated by the white arrows in Fig. \textbf{\ref{fig4r}b)}. 

The initial outward migration of C$_2$MImCl is fast (34.5 $\mu$m/s), while the subsequent charge reorganization revealed a diffusion-controlled exchange with two orders of magnitude slower exchange velocities; a similar result has been observed for LiCl (see Fig. \textbf{S4}), which is again consistent with weak ion-mica interactions. 
This also agrees well with the predictions of MD simulations \cite{kondrat2014accelerating}.

Our estimated diffusion coefficients in a molecular pore correlates with the diffusion coefficients reported in solids and along grain boundaries (in the range of 1$\times$10$^{-11}$ m$^2$/s) \cite{erdogdu2004determination}. 
This low diffusion coefficient is also consistent with the measured viscosity of electrolyte confined between mica and gold\cite{valtiner2012electrochemical}.

\begin{figure}[!t]
\includegraphics[scale=0.55]{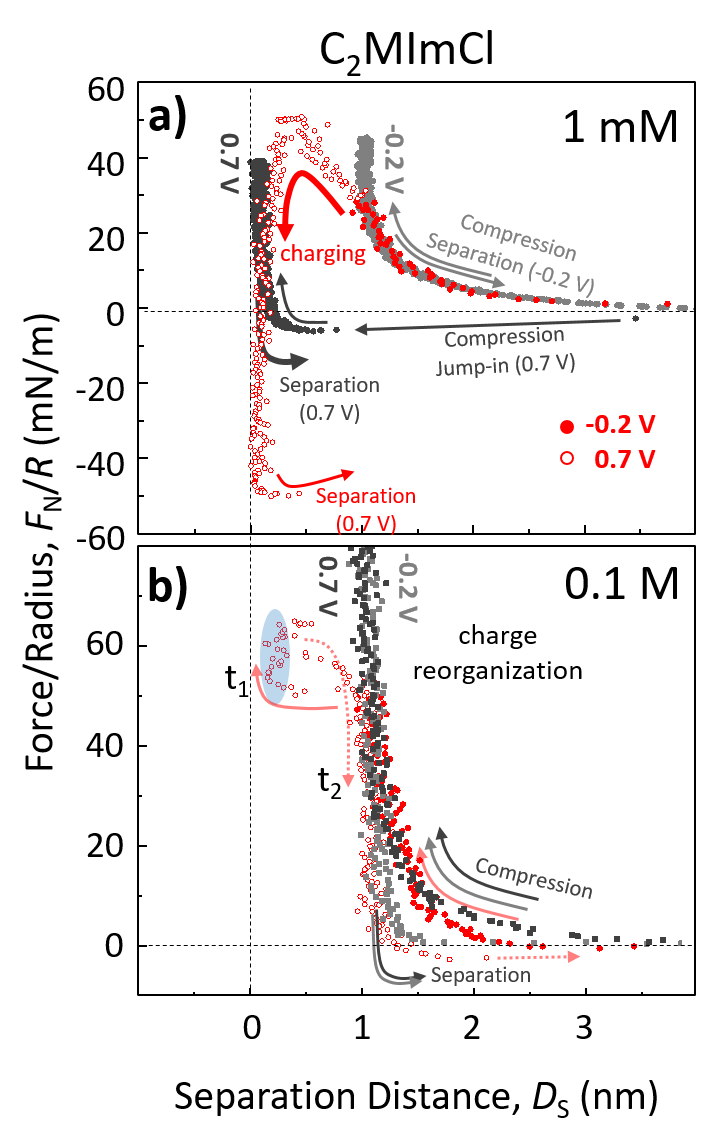}
   \caption{Dynamic and static sSFA F-D curves of \textbf{a)} 1 mM and \textbf{b)} 0.1 M C$_2$MImCl measured during polarization switch from -0.2~V (solid red circle) to +0.7~V (hollow red circle) and F-D profiles measured under constant gold surface polarization of -0.2~V (light gray) and +0.7~V (dark gray).}
   \label{fig6r}
\end{figure}

Getting back to the kinetically controlled two-step charge regulation, we can further access metastable interfacial structures in sSFA force measurements. 
First, we probed thermodynamically stable structures at every distance with the F-D measurements in equilibrium, \textit{i.e.} approaching very slowly from a large distance at constant applied potentials. Here, as expected from electrostatic interactions, we measure repulsive forces between negatively charged mica and negatively charged gold (at the applied potential of -0.2~V), and attractive forces at the positive gold polarization (+0.7~V).
We then extended F-D experiments to also detect non-equilibrated, kinetically trapped interfacial structures, where we switched the applied potential during the dynamic force measurements.
The measured F-D characteristics in 1 mM and 0.1 M C$_2$MImCl shown in Fig. \textbf{\ref{fig6r} a)} and \textbf{b)} compare equilibrated (fixed applied potentials of -0.2~V (light grey) and +0.7~V (dark grey)) and non-equilibrated system.

F-D characteristics before a switch in the gold polarization are plotted using solid red symbols and after the switch in hollow red.
Decompression was conducted shortly after the polarization change.
In this sense, we established an equilibrium structure during the approach but destabilized it at the turning point of the F-D measurement by inducing the electrochemical charge regulation.

First, in 1 mM solution, the polarization switch from -0.2~V to +0.7~V expelled the confined cations out of the gap. 
The surfaces then approached to $D_S$ $\approx$ 0, corresponding to a close contact as observed in the static-slit polarization experiments (see Fig. \textbf{\ref{fig4r}c)}). 
Interestingly, the measured adhesion was considerably larger, due to the enforced charge regulation. 

This indicates a strongly adsorbed ionic layer on the +0.7~V polarized gold surface, likely Cl$^-$, that can not be mechanically squeezed out by compression (black curve).
We observed the highest measured adhesion force for the case of compression at -0.2~V and a subsequent switch to +0.7~V. Polarizing to +0.7~V at the highest compression causes an electrostatically driven cation removal while the extremely confined volume kinetically prevented the rapid Cl$^-$ ion ingress. This allowed the measurement of a deeper van der Waals minimum.

At higher, 0.1 M concentration, we observed the same cation-expelling process, which shifted the surface separation close to D $\approx$ 0 (t$_1$ in Fig. \textbf{\ref{fig6r}b)}), but here, it displayed a clearly transient character.
The surfaces bounced in (t$_1$) and out (t$_2$) within a few seconds, indicating initial cation-depletion, followed by fast ion ingress (t$_2$, at +0.7~V). 
Hence, with this dynamic experiment we can further confirm that charge regulation in a stable molecular confinement can proceed \textit{via} metastable kinetic detours. 

In general, and as evidenced multiple times for various electrolytes, the competing migrative and diffusive flows can often result in multistep features that exposed the competition existing between the involved driving forces (with different timescales and energies) on the path to the (accessible) thermodynamic minimum.
Specifically, the subtle balance between enthalpic (ion-surface interaction) and entropic effects (ion depletion) at the slit opening determines the charge regulation mechanism in response to the external charging of a molecularly confined slit.

\section{Conclusions}
In summary, we experimentally demonstrated that the ion migration patterns can be well predicted from ion-surface interactions, and are easily accessed with molecular-resolution imaging techniques under non-confined conditions, which provides an experimental path to design charge regulation. 
Our data demonstrates that real dynamic systems may display multiple kinetic detours while achieving a thermodynamic equilibrium, as well as non-equilibrium meta-stable states.
We show that the occurrence of these meta-stable states can be tuned by adjusting ionic composition, concentration, or surface charge of the confining walls; something that can for example help to better control the highly confined electrokinetic flows in nanofluidic devices and facilitate their use for molecular sensing.
All in all, this is an exciting, potentially important, and hitherto unexpected aspect of the charge regulation in molecular confinement: multi-step paths, \textit{via} meta-stable conditions. 
These complex pathways may have important implications for all the mentioned processes, in which the confined charge regulation plays a critical role, and we speculate that especially in biological systems this may not just be a feature but also a functionality

\section{Materials and Methods}
\paragraph{Chemicals}
Ionic liquid aqueous solutions with various concentrations were prepared by mixing \textgreater 97 \%, pure 1-Methyl-3-octylimidazolium chloride (C$_8$MImCl, Alfa Aesar) or 1-Ethyl-3-methylimidazolium chloride (C$_2$MImCl, purity: \textgreater 98 \%, Alfa Aesar) with Milli-Q water (organic impurities: $\sim$ 2\ ppb). Ionic liquid concentration series were prepared from 2 M stock solutions.

\paragraph{Materials}
Back-silvered, molecularly-smooth muscovite (001) mica sheets (optical grade V1), supplied by S\&J Trading Company; USA, were used as substrates in all sSFA and AFM experiments. 
For mica-gold experiments, we used mica template-stripped gold films (purity 99.99\%) with the rms roughness < 0.5 nm. 

\paragraph{In-situ sensing Surface Force Apparatus}
In this work we use a real-time sensing surface force apparatus (sSFA) more thoroughly described in our recent work.\cite{Wieser2021} 
This instrument uses the same interferometrical principle as SFA/SFB\cite{israelachvili2010recent} to measure force-distance characteristics between two mirrors. 
Analysis of interferometrical patterns to obtain separation distance is done using a transfer matrix method.\cite{schwenzfeier2019optimizing}
In addition, the sSFA combines the advantages of the traditional, high precision distance measurement \textit{via} multiple beam interferometry with a newly-introduced pair of customised force sensors with $\mu$N sensitivity (load-cell), which are central for being able to independently measure the force and the separation distance (unlike in the traditional SFA approach, where forces are calculated from the measured distance change of the center of the contact, see explanation in Fig. \textbf{S5}. 

With the independent force measurement we are able to use the live force readout to establish the thermal drift while the surfaces are still separated.
We then placed the surfaces into a position with a separation distance of a few nanometers, at which the surfaces were forming a flattened contact region (see FECO shape in Fig. \textbf{\ref{fig1r}c)}). This required the applied loads ranging between 0.0 and 5.0 mN/m, depending on the height of the repulsive EDL barrier in a given solution.
We calculated the mechanical drift from the difference between the force slopes at large separation distances (50-100 nm) and separations of a few nanometers. This difference arises from the addition of the mechanical drift to the thermal drift. 
Once the linear slope of the mechanical drift was established, a counter motion with the piezo could be applied to compensate it. 
The drift compensation keeps the surfaces at a mechanical drift-free distance while allowing the system to freely adjust to a new equilibrium distance induced by the polarization as shown in Fig. \textbf{\ref{fig4r}c)} and \textbf{\ref{fig5r}}. 
\paragraph{Surface Fabrication and Electrochemistry}
All electrochemical experiments were performed with a 3 electrode setup using a PalmSense 4 or a Biologic Potentiostat. 
In sSFA measurements, back-silvered thin mica sheets (5 - 10 $\mu$m) were glued to cylindrical glass disks with a radius of curvature of $\sim$ 2 cm. 
The working electrode consists of 40 nm thick gold substrate glued on another cylindrical disk via template stripping and is connected with the potentiostat using a thin teflon-coated gold wire (Goodfellow). 
A standard Ag|AgCl mini electrode comprises the reference electrode, and the counter electrode is a circular platinum wire or a mesh (Goodfellow), placed around the working electrode (see Fig. \textbf{\ref{fig1r}a)}) 

\paragraph{Procedure for sSFA force measurements}
In our force measurements, we only used mica-mica or mica-gold contacts that fulfilled the following cleanliness criteria: The chosen contacts between the surfaces show the expected and reproducible \textbf{1)} attractive jump-in during approach and \textbf{2)} adhesive jump-out in initial measurements in Milli-Q water. 
Therefore, the reference Milli-Q water measurement was always conducted before introducing aqueous ionic liquid mixtures or chloride salt solutions for each used surface pair. (see Fig. \textbf{S2} and \textbf{S6})
For experiments in molecular confinement the surfaces were placed to a separation distance of a few nanometer via the piezo. 
Due to the drift compensation the separation distance change is attributed to re-equilibration without influence of thermal or mechanical artefacts. 
Force-distance curves were recorded with the sampling rate of spectrometer and force sensor of 10 Hz.
Distance data points calculated with the SFA explorer\cite{schwenzfeier2019optimizing} were then correlated with the measured real-time force values. 

\paragraph{Blue-Drive Atomic Force Microscopy}
All shown AFM topography images and phase-distance profiles were acquired using a Cypher S atomic force microscopy system (Asylum Research) in amplitude modulation (tapping) mode. 
AFM data analysis was performed in AR based on Igor Pro 6.3 and Gwyddion 2.55. 
Further data analysis was performed in Python 3.7.
Reflex gold coated ultra high frequency silica probes (ARROW-UHFAuD from NanoWorld) with a resonance frequency of 0.7 - 2.0 MHz in air were employed for all AFM measurements. 
The cantilever oscillation was driven by blueDrive photothermal excitation. 
For high resolution imaging the laser power was set between 2.7 and 3.0 mW and images were recorded with a resolution of 256 points and lines at a rate of 4 to 8 lines per second, and imaging set points were adjusted to reveal ion adsorption layers (see again Fig. \textbf{S1}.)

\paragraph{Diffusion coefficient estimation}
We estimate the characteristic diffusion coefficient (D$_C$) for ion pairs in molecular confinement, using the data marked on Figure \textbf{\ref{fig4r}b}.
D$_C$ is calculated using a simplified 2D radial Fickian diffusion model, where the diffusion distance \textit{x} can be estimated as follows\cite{mukhopadhyay2002contrasting}:

\begin{equation}
x^2 = 4D_Ct    
\end{equation}

with the diffusion coefficient \textit{D$_C$} and time \textit{t}.

\section{Data availability}
The data that support the findings of this study are available on reasonable request from the corresponding author, M.V. and H.-W.C..

\section{Acknowledgements}
The authors acknowledge support from the European Research Council (Grant: CSI.interface, ERC-StG 677663, characterization of EDLs).
J.D. acknowledges support from the Research Council of Norway, FRIPRO grant nr. 286733. 
The authors thank Yana de Smet for the help with AFM imaging.

\section{References}
\bibliography{References.bib}
\bibliographystyle{ieeetr}

\end{document}

\title{Supplemental Information for:\\*
Real-time visualization of metastable charge regulation pathways in molecularly confined slit geometries} 

\author{H.-W. Cheng}
\email{hsiu-wei@iap.tuwien.ac.at}
\affiliation{Institute of Applied Physics, Vienna University of Technology}

\author{J. Dziadkowiec}
\affiliation{Institute of Applied Physics, Vienna University of Technology}
\affiliation{NJORD Centre, University of Oslo}
\author{V. Wieser}
\affiliation{Institute of Applied Physics, Vienna University of Technology}

\author{A. M. Imre}
\affiliation{Institute of Applied Physics, Vienna University of Technology}

\author{M. Valtiner}
\email{valtiner@iap.tuwien.ac.at}
\affiliation{Institute of Applied Physics, Vienna University of Technology}

\maketitle

\begin{figure}[!htb]
    \centering
    \includegraphics[scale=0.1]{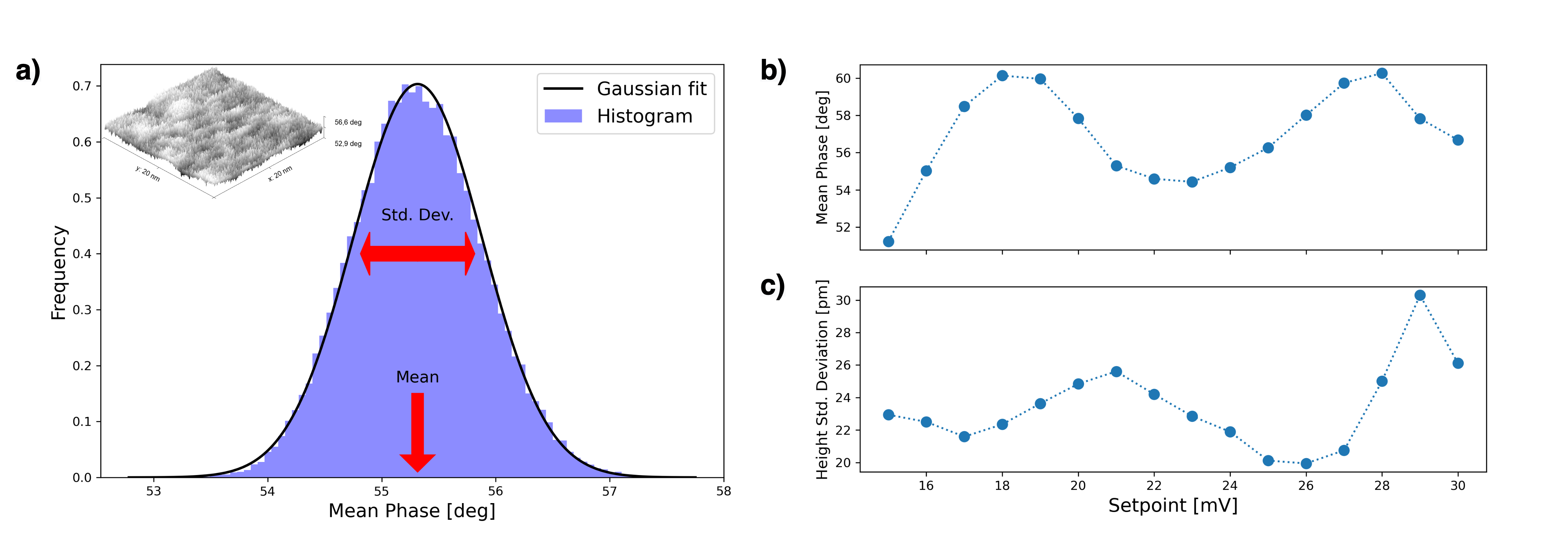}
    \caption{Summary of additional quantitative analysis of AFM topographies obtained for 2 M C$_8$MImCl on mica substrate. Resolving the surface lattice requires a low setpoint, i.e. pressing hard on the surface. When the setpoint is increased, the image first becomes noisy and upon further increase the outer layer is resolved. Images were recorded for a number of setpoints around this point. \textbf{a)} Height and phase values for the images are transformed into a histogram and fit to a Gaussian distribution giving a mean and a standard deviation value. The mean phase shift (\textbf{b)}) and the standard deviation of the height values (\textbf{c)}) are plotted against the used setpoint. This reveals an oscillatory change over the setpoint variation, which agrees well with the AFM phase-distance profile and sSFA Force-Distance curves shown in Fig. \textbf{2)}.}
    \label{Supfig1r}
\end{figure}

\begin{figure}[!htb]
    \centering
    \includegraphics[scale=0.4]{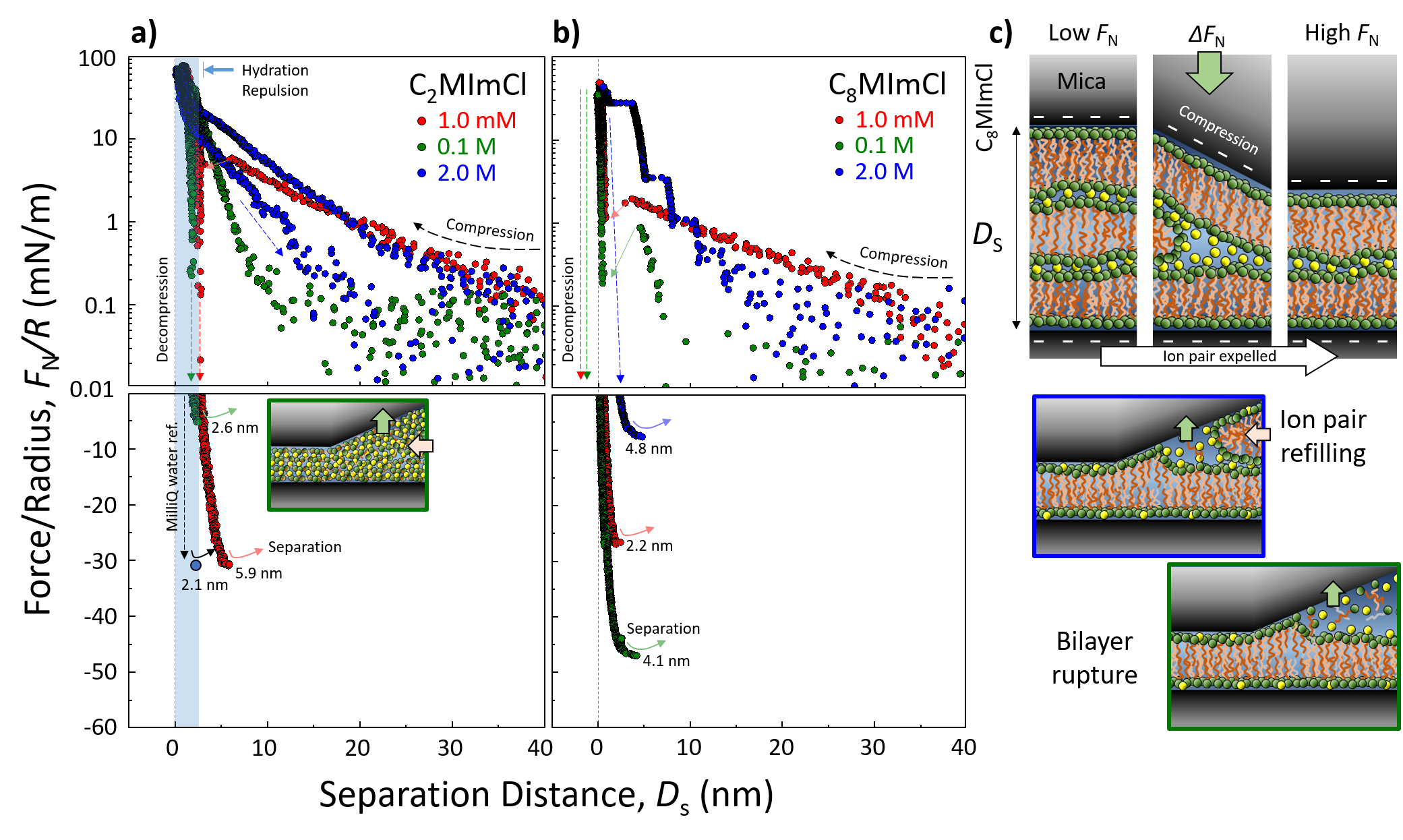}
    \caption{sSFA force-distance profiles measured between two apposing mica surfaces in 1 mM, 0.1 M and 2 M of \textbf{a)} C$_2$MImCl and \textbf{b)} C$_8$MImCl aqueous solutions. Repulsive forces are plotted on a logarithmic scale (top panels) and adhesive forces (F\textsubscript{N}/R< 0) on a linear scale (bottom panels). Only the data preceding the adhesive jump-out is shown for the retraction profiles. The reference adhesion force measured in MilliQ water is marked as single point, corresponding to the last data point before the adhesive jump-out in a given system. The forces measured in 1 mM (red) \textbf{a)} C$_2$MImCl and \textbf{b)} C$_8$MImCl solutions agree with the theoretical DLVO interactions, typical for inorganic electrolytes (with fitted Debye lengths; \(\lambda\)\textsubscript{D} of 9.52 nm, 0.9 \% error; and 9.80 nm, 2.0 \% error, respectively; theoretical \(\lambda\)\textsubscript{D} = 9.61 nm).
    A very clear multilayered structure at mica/IL interface with an individual layer thickness of \(\sim\)2.2 nm was observed. The characteristic jump-in distances for each layer are within a narrow range of \(2.20\pm0.08\) nm, which is indicative of the consecutive expulsion of stable, confined ion layers. Each confined IL layer shows a pronounced compression/expansion of about 30 \% before it is expelled and the surfaces jump into closer confinement or apart, respectively. Such behaviour suggests that flexible lamellar structures may adapt to the positive and negative loads at a molecular level, by damping or enhancing the thermal fluctuations, or \textit{via} molecular tilting. The measured adhesive minima increase exponentially with the decreasing number of confined layers (Figure \textbf{\ref{Supfig3r}a)}. This is not the expected power law relationship that is characteristic for a purely van der Waals distance-dependent adhesion of contacting macroscopic bodies and suggests a strong ion-ion correlation forces that mediates each adhesive minimum. 
    Sketches in panel \textbf{c)} illustrate the proposed compression and color-coded separation mechanisms for C$_8$MImCl, where F$_N$ is the applied normal force.}
    \label{Supfig2r}
\end{figure}

\begin{figure}[!htb]
    \centering
    \includegraphics[scale=0.7]{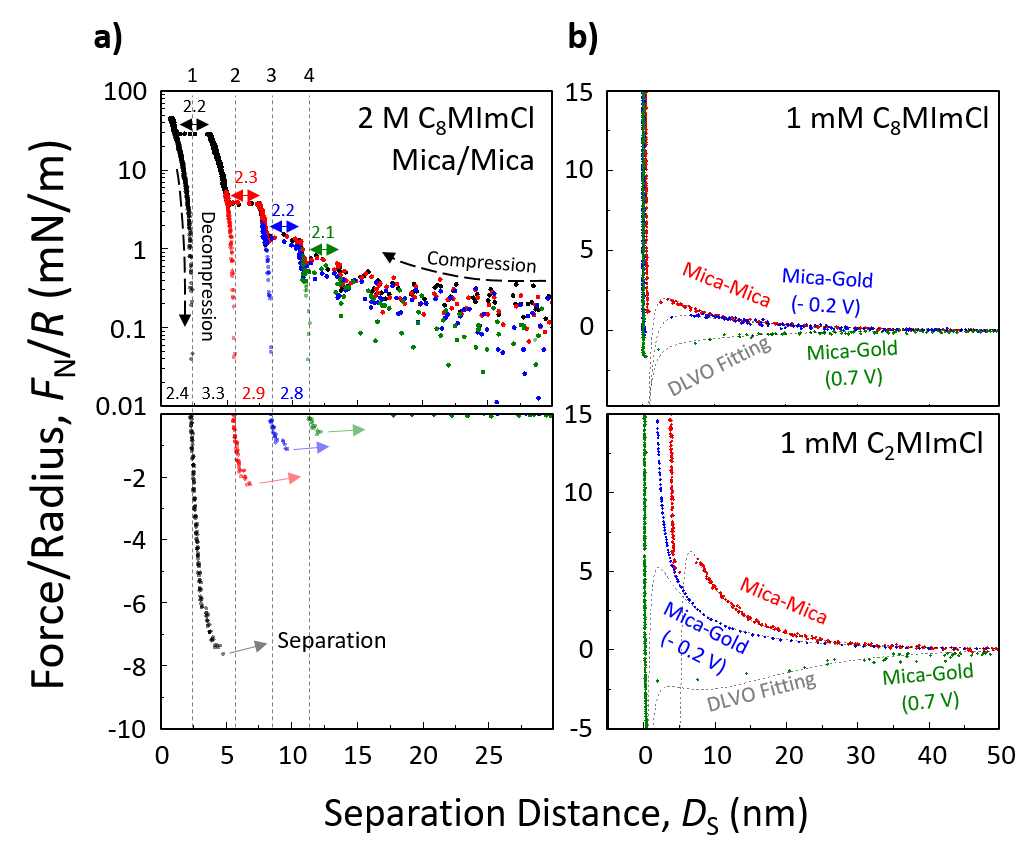}
    \caption{\textbf{a)} sSFA force-distance (F-D) profiles with the consecutively increasing maximum applied load in 2 M C$_8$MImCl solution, confined between two apposing mica surfaces. \textbf{b)} DLVO fitted F-D profile of 1 mM C$_2$MImCl (bottom) and C$_8$MImCl (top) measured between two mica surfaces and in electrochemically modulated mica-gold configuration of -0.2~V and +0.7~V. At low concentrations, IL electrolytes behave as classic electrolytes, and data can be fitted well with a charge-regulation model.}
    \label{Supfig3r}
\end{figure}

\section*{Charge-regulated DLVO Fitting}
 The following charge-regulation formula adapted from Bilotto \textit{et. al.}\cite{bilotto2019interaction} is used to fit DLVO parameters in this work:

\begin{multline}
\frac{F_{DLVO}}{R}\, = \frac{F_{VDW}}{R}\, + \frac{F_{EDL}}{R}\, = -\frac{A}{6(D-D_{VdW})^2}\, \\ \,
+ \, 2\pi \epsilon \epsilon_0 \kappa \frac{\left[2\,\psi_1 \psi_2 \, e^{-\kappa \,(D-D_{EDL})}+ \left( \left( 2 \, p_1 \, -1\right)\, \psi_2^2 \, + \left( 2 \, p_2\, -1\right)\, \psi_1^2 \right)\, e^{-2 \,\kappa\,\left(D - D_{EDL}\right)}\right]}{1-\left(2p_1 - 1 \right)\left(2 p_2 -1\right) \, e^{-2 \,\kappa\,\left(D - D_{EDL}\right)}}
\label{equation1}
\end{multline}

\begin{table}[!htb]
\centering
\begin{tabular}{|m{1.8cm}|m{2.2cm}|m{5cm}|m{4cm}|}
    \hline
    IL System & Debye Length 1/$\kappa$ (nm) & Gold/Mica Charge-Regulation Parameters $P_1/P_2$ (-) & Gold/Mica Surface Potential $\psi_1/\psi_2$ (mV) \\
    \hline
    C$_2$MImCl (-0.2~V) & 9.52 & 1/0.78 & -70/-60\\
    \hline
    C$_2$MImCl (0.7 V) & 9.52 & 0/0.78 & +24/-60\\
    \hline
    C$_8$MImCl (-0.2~V) & 9.8 & 0.2/0.77 & -53/-52\\
    \hline
    C$_8$MImCl (0.7 V) & 9.8 & 0.85/0.77 & +35/-52 \\
    \hline
    \end{tabular}
\label{table1}
\caption{Parameters and DLVO fits for 1 mM C$_2$MImCl and C$_8$MImCl aqueous solutions under applied surface polarization of -0.2~V or 0.7 V}
\end{table}

\begin{figure}[!htb]
    \centering
    \includegraphics[scale=0.45]{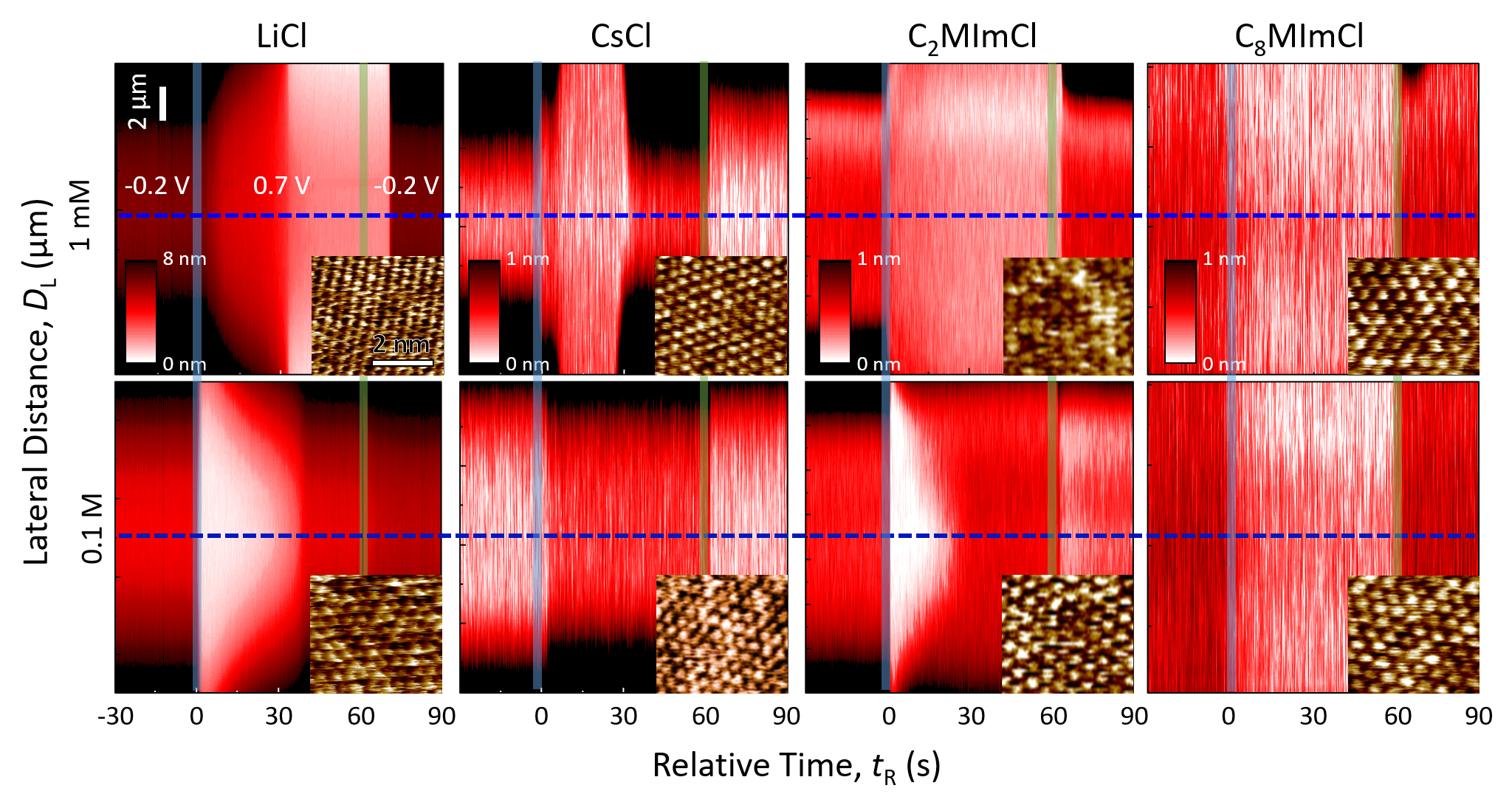}
    \caption{Time-resolved, 3D sSFA lateral slit thickness profiles for 1 mM and 0.1 M LiCl, CsCl, C$_2$MImCl and C$_8$MImCl aqueous solutions in the mica-gold configuration under the applied square-wave polarization oscillating between -0.2~V and +0.7~V.
    The color bar coding from white to black corresponds to separation distance D$_S$ (mica/gold slit thickness) ranging from 0 to 8 nm for LiCl, and 0 to 1 nm for the other salts. The 3D resolved sSFA separation distance maps show the response of the gap opening and closing, while it reacts to the polarization-induced ion saturation or ion depletion similar to the maps shown in Figure \textbf{4}. The data corresponds to the separation distance centre line profiles shown in Figures \textbf{4c)} and \textbf{5}. 
    The insets show AFM images of the inner EDL structure at the mica-solution interface and relate the observed ion migration behaviour to the strength of ion-mica interactions, with poorly or strongly ordered cation layers, obtained as described in the Materials and Methods section.}
    \label{Supfig4r}
\end{figure}

\begin{figure}
    \centering
    \includegraphics[scale=0.7]{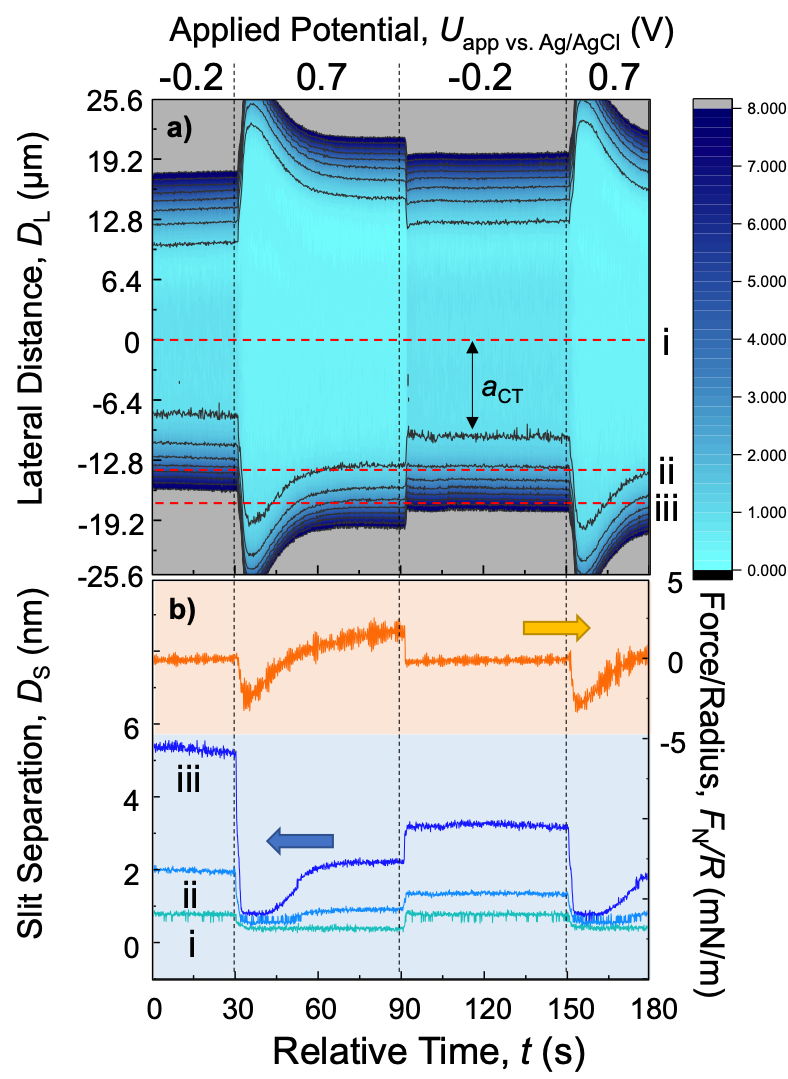}
    \caption{sSFA slit polarization experiment in 10 mM C\textsubscript{2}MImCl solution corresponding to those shown in Figure \textbf{S4}. Here, we extract the slit separation distances at various lateral distances from the centre of confinement between mica and gold surfaces and show the corresponding change in the total measured force (plotted in orange). The applied potential is continuously shifted between -0.2~V and +0.7~V as in all shown slit-charging experiments. \textbf {a)} Cross section of the variation in geometry of the flattened contact region between mica and gold surfaces  plotted against time. The color bar shows the slit separation at a given lateral distance \textit{D}\textsubscript{L}, where the color code from light blue to dark blue represents the slit separation from 0 nm to 8 nm, respectively, with scaling increments of 1 nm for each color. The grey region represents slit separations greater than 8 nm. The change of contact radius a$_{CT}$ is indicated by the first grey line. \textbf{b)} Slit separation against time at three selected lateral positions where \textit{D}\textsubscript{L} = 0 (center of slit), \textit{D}\textsubscript{L} = 13 $\mu$m and \textit{D}\textsubscript{L} = 18 $\mu$m from the center. The measured total normal force change over the same time is plotted in orange.}
    \label{Supfig5}
\end{figure}

\begin{figure}
    \centering
    \includegraphics{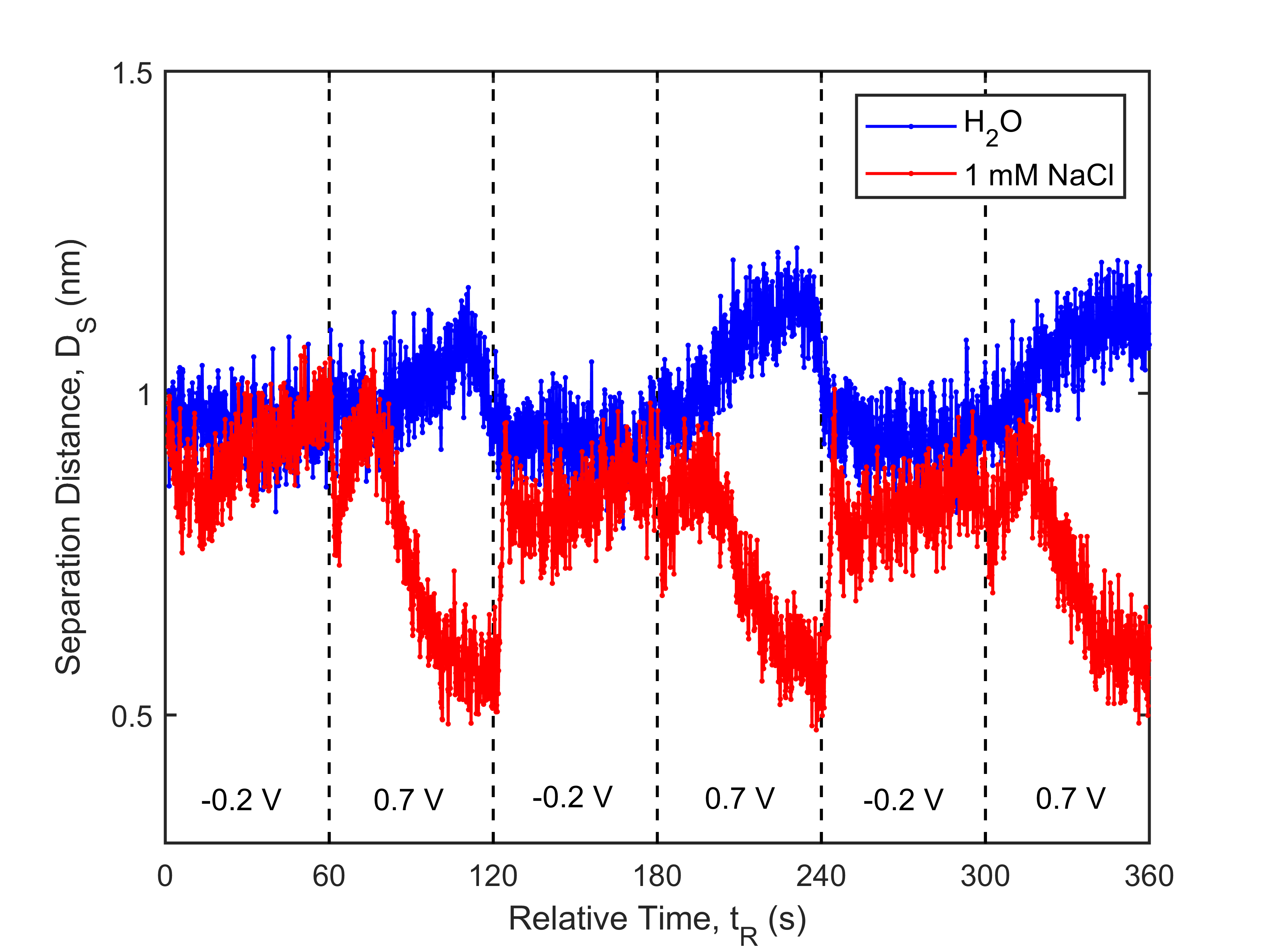}
    \caption{Comparison of separation distance between mica and gold surfaces (measured in the center of confinement in the sSFA) \textit{vs.} time upon the change in the applied potential to the gold surface between -0.2~V and +0.7~V for MilliQ water (blue) and  1 mM NaCl solution (red). The presented data for water and 1 mM NaCl solutions were measured in the same experiment (with the same pair of mica and gold surfaces and in the same contact region). The 1 mM NaCl line profile corresponds to the one shown in Figure \textbf{5}.} 
    \label{Supfig6}
\end{figure}

\clearpage
\bibliography{References.bib}
\bibliographystyle{ieeetr}